   \documentclass[prl,aps,twocolumn,showpacs]{revtex4}
\usepackage[dvips]{graphicx}
\usepackage{dcolumn}
\newcommand{\be}{\begin{equation}}
\newcommand{\ee}{\end{equation}}
\newcommand{\bea}{\begin{eqnarray}}
\newcommand{\eea}{\end{eqnarray}}
\newcommand{\nn}{\nonumber}

\begin{document}
\topmargin=-20mm
%     \Large

\title{Theory of Magneto-Acoustic Transport in Modulated  Quantum Hall Systems Near $\nu=1/2$}

\author{\large Nataliya A. Zimbovskaya and Joseph L. Birman}

\address{Department of Physics, The City College of CUNY, New York, NY, 10031, USA}

\date{\today}

\begin{abstract}   
Motivated by the experimental results of Willett et al [Phys.Rev. Lett. {\bf 78}, 4478 (1997)] we develop a magnetotransport theory for the response of a two-dimensional electron gas (2DEG) in the fractional quantum Hall regime near Landau level filling factor $ \nu = 1/2 $ to the surface acoustic
wave (SAW) in the presence of an added periodic density modulation. We assume there exists a composite fermion Fermi surface (CF-FS) at $ \nu = 1/2 $, and we show that the deformation of the circular CF-FS due to the density modulation can be at the origin of the observed transport anomalies for the experimental conditions. Our analysis is carried out particularly for the nonlocal case which corresponds to the
SAW experiments. We introduce a new model of a deformed CF-FS.
The model permits us to explain anomalous features of the response of the modulated 2DEG to the SAW near $ \nu = 1/2: $ namely, the nonlinear wave vector dependence of the electron conductivity, the appearance of peaks in the SAW velocity shift and attenuation and the anisotropy of the effect, all of which originate from contributions to the conductivity tensor due to the regions of the CF-FS, which are flattened by the
applied modulation.
              \end{abstract}

\maketitle

\section{I. Introduction}

Surface acoustic waves (SAW) propagating above a two dimensional
electron gas (2DEG)  in GaAS/AlGaAs heterostructures in a strong
magnetic field change their properties due to the conductivity of the
adjacent electron system.  Hence, the SAW experiments provide an efficient tool
for experimental studies of the conductivity of two dimensional
electron systems in Quantum Hall states at even denominator filling
factors $ \nu $ \cite{1,2,3,4,5,6}. It was observed repeatedly that the dependence
of the SAW velocity shift on the magnetic field $ B $ exhibits a
pronounced minimum at $ B $ corresponding to $ \nu = 1/2, $ implying a maximum in the conductivity. Exactly at $ \nu = 1/2 $ the
conductivity appeared to be linear in the sound wave vector $ q $ when the wave length is small enough. These results
are in good agreement with theoretical calculations based on the
Chern-Simons approach of Halperin, Lee and Read (HLR) \cite{7,8}, describing the Quantum Hall State at and near $ \nu = 1/2, $ thus providing strong support for this theory. The HLR theory corresponds to the physical picture of the electrons decorated by attached quantum
flux tubes which are the relevant fermionic quasiparticles of the
system (so called composite fermions (CF)). Near half filling the CFs
move in the reduced effective magnetic field $ B_{eff} = B - 4 \pi
\hbar c n/e \ (n $ is the electron density). Just at $ \nu = 1/2 $
the CFs form a Fermi sea and exhibit a Fermi surface (FS). The CF-FS
is a circle in a two dimensional (2D) quasimomenta space and its
radius $ p_F $ equals $ \sqrt{4 \pi n \hbar^2}. $

Due to the piezoelectric properties of GaAS, the velocity shift
$ (\Delta s/s) $ and the attenuation rate $ (\Gamma) $ for the SAW,
propagating along the $ x $ axis across the surface of a
heterostructure, containing 2DEG, take the form \cite{9,10,11}:

  % f 1-2
                    \bea &&
\frac{\Delta s}{s} = \frac{\alpha^2}{2} \mbox{Re}
\left( 1 + \frac{i \sigma_{xx}}{\sigma_m} \right )^{-1} ;
                  \\ \nn \\ &&
\Gamma = -q \frac{\alpha^2}{2} \mbox{Im} \left( 1 + \frac{i
\sigma_{xx}}{\sigma_m} \right )^{-1} .
                   \eea
 Here $ {\bf q},\omega = s q $ is the SAW wave vector and frequency,
$ \alpha $ is the piezoelectric coupling constant, $ \sigma_m =
\varepsilon s/2 \pi, \; \varepsilon $ is an effective dielectric
constant of the background and $ \sigma_{xx} $ is the component of
the electron
conductivity tensor.

According to the semiclassical CF theory, the electron resistivity
tensor $ \rho $ (at finite $ q $ and $ \omega) $ is given by \cite{8}:
% f 3
                \begin{equation}
                \rho = \sigma^{-1} = \rho^{CF}  + \rho^{CS}
                                  \end{equation}
 where $ \rho^{CF} $ is the CF resistivity tensor which has to be
calculated by means of a Boltzmann transport equation; the
contribution $\rho^{cs}$ arises due to a fictitious magnetic field
which originates from the Chern-Simons formulation of the theory.
This tensor containes only off diagonal elements $ \rho_{xy}^{CS} = -
\rho_{yx}^{CS} = 4 \pi \hbar /e^2. $

This work was stimulated by the new results obtained near
$ \nu = 1/2 $ on samples whose electron density was periodically
modulated in one direction by applying an additional static electric field \cite{12,13}. In summary, it was observed in the experiments \cite{12} that the periodic
density modulation in the 2DEG produces dramatic and highly
anisotropic response to the SAW. When the density modulation wave
vector {\bf g} is orthogonal to the SAW wave vector {\bf q}, the
minimum in the magnetic field dependence of the sound velocity shift at $\nu = 1/2 $ was converted to a large maximum. The maximum was
observed only for sufficiently large modulation wave vectors and
sufficiently high magnitudes of the electric field producing the
electron density modulation. In some experiments reported in Ref. \cite{12} the
peak disappeared on further increasing  the electron density
modulation but was replaced by a minimum in the SAW velocity shift
again. In the case when the SAW propagated along the electron density
modulation direction no anomaly in the electron system response was
observed.

In the framework of the HLR approach we can see that the grating will influence the CF system in two ways: through the direct effect of the modulating potential and through the effect of the magnetic field $ \Delta B({\bf r}) $ proportional to the local density modulation $ \Delta n({\bf r}) \ [\Delta B({\bf r}) = - 4 \pi \hbar c \Delta n({\bf r})/e ].$ The latter
was analysed recently \cite{14} under the conditions $ q << g \; ql << 1 \ (l $ is the mean free path of the CF). It was shown that under these conditions the corresponding component of the 2DEG conductivity $(\sigma_{xx}, $ for the SAW propagating along $ x $ axis at $ {\bf q}
\perp {\bf g}) $ has an additional term proportional to $ (\Delta
n/n)^2 \ [\Delta n $ is the amplitude of the density modulation $
\Delta n ({\bf r})]. $ This term arises due to the inhomogeneity of
the field $ B_{eff} $ in modulated structures and increases on 
increasing $ B_{eff} .$ According to Eq. (1), it corresponds to the
peak in the SAW velocity shift $ \Delta s /s $ at $\nu = 1/2. $
Similar results under the same conditions $ (q << g, \; ql << 1) $
were obtained in Ref. \cite{15}.

The relative magnitude of the effect of electric modulations compared to magnetic modulations was estimated in Refs. \cite{16,17,18} in which the magnetotransport problem in modulated 2DEG in low magnetic fields was treated. It was shown that corrections to the conductivity arising due to  magnetic modulations are larger by a factor $ (k_F/g)^2 $ than the corresponding corrections in the case of the electric modulations for equal modulation strength. 
In recent work by two groups \cite{14,15}, this estimate was applied to modulated 2DEG in quantum Hall regime, and as a result, they concluded that the effect of electric modulating potential is small compared to the effect of inhomogeneity of the effective magnetic field.

However, neither Refs \cite{16,17,18} nor \cite{14,15} took into account that a periodic modulating electric field can deform the originally circular CF-FS. We can recall the analogous situation of the action of  an extra crystalline field on the Fermi surface of a conventional metal. In the case of a modulated 2DEG, the modulation wave vector $ \bf g $ replaces the reciprocal-lattice vector of the conventional metal.

In the {\it local} regime this deformation of the CF-FS does  not significantly influence results concerning dc transport of  the modulated 2DEG both in quantum Hall regime and in low magnetic fields, because under these local conditions all segments of the FS of relevant quasiparticles contribute to the response functions essentially equally. Thus the difference in shape between the deformed and undeformed FSs does not lead to any significant changes of the 2DEG conductivity. So in study of dc transport problems, the deformation of the FS by a modulating potential can be ignored, and the effect of both electric and magnetic modulations can be treated as small perturbations, which cause the appearance of small corrections to the Drude cconductivity tensor of the corresponding homogeneous CF system.

In contract, the response of a modulated 2DEG to an extra perturbation (in our case SAW) propagating in the {\it nonlocal} regime can be essentially changed due to a small deformation of the CF-SF. For the nonlocal regime $ (ql >1) ,$  the most important mechanism of absorption of the energy of the electric field accompanying the SAW is not connected with quasiparticle scattering: it is Landau damping. Therefore, the main contribution to the conductivity originates from those small effective parts of the FS where the quasiparticle velocity vector $ \bf v $ and the SAW wave vector $ \bf q $ are nearly transverse $ \bf ( q \cdot v ) \approx $0.

Again we recall that in metals the local geometry of the  "effective" parts of the FS can
influence significantly  the electron conductivity and, consequently,
the acoustic attenuation and the velocity shift in the limit
of large $ q \  (ql >> 1)$ \cite{16,17,18}. The conductivity and associated
observables are extremely sensitive to small local anomalies in the FS
geometry such as parabolic points or points of flatness and to a local
change in the topology of the FS. The linear response functions of the
CFs at $ ql > 1 $ are also expected to be sensitive to the local changes of their
FS geometry.

In the study of nonlocal response of modulated CF system, we cannot neglect the effect of deformation of the CF-FS. As a cause of the deformation of the CF-FS, the scalar potential modulation caused by the ruled grating can be very important at $ ql > 1 $ although the change of CF energy due to the electric modulation is small. In this paper we will show that a
modulation-induced deformation of the CF-FS can be at the origin of the observed transport anomalies.

\section{II. The model}

The CF-FS is
assumed to be a circle at $ \nu = 1/2, $ in the absence of modulation.
It can be significantly distorted  due to the modulating electric field
as we now show. Assume that electronic density is modulated
in the  the $ y$-direction and the modulation period is small enough
$ (\hbar g > 2 p_F). $ Using the nearly free electron model one
can obtain the energy-momentum relation for the CF in the form:
% f 4
               \begin{equation}
E({\bf p}) = \frac{p_x^2}{2 m^*} +  \frac{p_y^{*2}}{2 m^*} +
\frac{(\hbar g)^2}{8 m^*} -
\sqrt {\left (\frac{\hbar g p_y^*}{2 m^*} \right)^2 + V_g^2} \, .
                   \end{equation}
Here $ p_y^* = p_y - \hbar g/2, \;m^* $ is the CF effective mass;
$ V_g $ is the magnitude of the quasiparticle potential energy
in the periodic electric field.
$ V_{g} $  is assumed to be small in comparison with the quasiparticles Fermi energy
$ E_F. $ Calculating the FS curvature: 
% f 8
               \begin{equation}
K = \frac{1}{v^3} \left (2 v_x v_y \frac{\partial v_x}{\partial p_y} -
v_x^2 \frac{\partial v_y}{\partial p_y} -
v_y^2 \frac{\partial v_x}{\partial p_x} \right )
                   \end{equation}
with $ v = \sqrt{v_x^2 + v_y^2} $ one can find it tending to zero when
$p_x $ tends to $ \pm p_F (V_g/E_F)^{1/2} $ (See Fig.1).

\begin{figure}[t]
\begin{center}
\includegraphics[width=5.4cm,height=6cm]{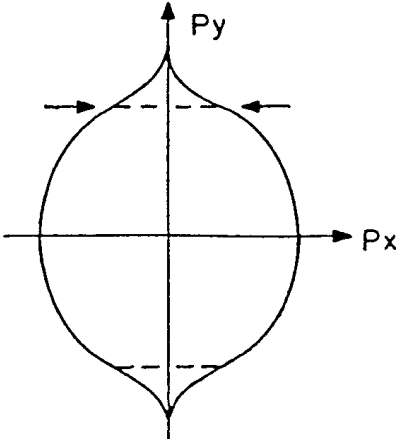}
\caption{The shape of the CF-FS deformed due to an external periodic
modulation in the ''y'' direction in the almost free electron approximation
(solid line) at $ 2 p_F < \hbar g $ and the CF-FS corresponding to
Eq.(8) (dashed line). }
\label{rateI}
\end{center}
\end{figure}

In the vicinities of the corresponding points on the FS the
quasiparticles velocities are nearly   parallel to the
$y$-direction. So these parts of the FS contribute strongly to
the attenuation and velocity shift of the SAW,  propagating in
the $x$-direction.
Near these zero curvature points we will
use asymptotic expressions for Eq.(4). Determining
$(p_{x_0}, p_{y_0})$ by $ p_{x_0} = \zeta p_{F} $, $ \displaystyle
{p_{y_0} = p_{F} \left [1-(1/\sqrt2) \zeta^{2} \right ]} $, where
$\zeta=\sqrt{V_{g}/E_{F}}$, $E_{F}=p_{F}^{2}/2m^{*}$, we can expand the
variable $p_{y}$ in powers of $(p_{x} - p_{x_0}), $ and keep the lowest-order terms in the expansion. We obtain:
  % f6
       \begin{equation}
 p_{y}-p_{y_0} = -\zeta(p_{x} - p_{x_0}) -
  \frac{2}{\zeta^{4}}\frac{(p_{x} - p_{x_0})^{3}}{p_{F}^{2}}.
                    \end{equation}
 Near $p_{x0}$, where $(|p_{x} - p_{x_0}| < \zeta^{2}p_{F}) $ the first term on the right side of Eq.(6) is small compared to the second one and can be omitted.

\begin{figure}[t]
\begin{center}
\includegraphics[width=5.4cm,height=6cm]{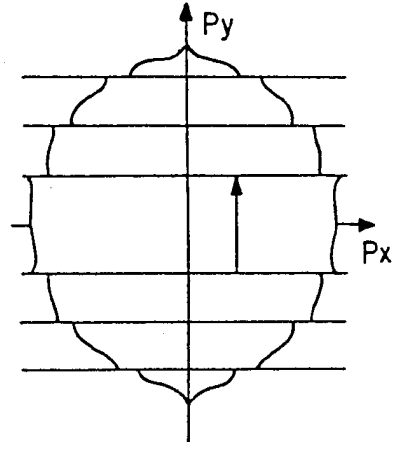}
\caption{ The deformation of the CF-FS in modulated 2DEG system in the almost
free electron approximation under the condition  $ 2 p_F < \hbar g .$ The
modulation wave vector {\bf g} is directed along the ''y'' axis. }
\label{rateI}
\end{center}
\end{figure}

Hence near $p_{x_0} $ we have:
% f7
 \begin{equation}
E({\bf p})=\frac{4}{\zeta^{4}}\frac{p_{F}^{2}}{2m^{*}}
\left(\frac{p_{x}-p_{x_0}}{p_{F}}\right)^{3}+\frac{p_{y}^{2}}{2m^{*}}.
          \end{equation}
The ''nearly free'' particle model can be used when $ \zeta^2 $ is very
small.
For larger $ V_g $ corresponding to $ \Delta n/n $ of the order of a few
percent (as in the experiments [12]) the local flattening of the CF-FS can be
more significant.
To analyze the contribution to the conductivity from these flattened
parts we  generalize the expression for $E({\bf p})$ and define our
dispersion as:
% f8
\begin{equation}
E({\bf p}) = \frac{p_{0}^{2}}{2m_{1}} \left |
\frac{p_{x}}{p_{0}} \right |^{\gamma} + \frac{p_{y}^{2}}{2m_{2}},
             \end{equation}
where $p_{0}$ is a constant with the dimension of momentum, the
$m_{i}$ are effective masses, and $\gamma$ is a dimensionless
parameter which will determine the shape of the CF-FS .
When $\gamma>2$ the 2D CF-FS looks like an ellipse
flattened near the vertices $(0, \pm \sqrt{m_2/m_1} p_{0})$.
Near these points the curvature is:
% f9
\begin{equation}
K = - \frac{\gamma(\gamma-1)}{2p_0 \sqrt{m_1/m_2}}
\left| \frac{p_x}{p_0} \right|^{\gamma-2}
\end{equation}
and, $ K \rightarrow 0$ at $p_{x} \rightarrow 0 $. The CF-FS will be
the flatter at $ (0, \pm \sqrt {m_2/m_1} p_0)$,
the larger is the parameter $ \gamma $. A separate investigation is required
to establish how $ \gamma $ depends on $ V_g.$ Here we postulate Eq.(8) as
a natural generalization of Eq.(7) and we then derive the resulting SAW
response. Thus (8) is our concrete new model for the deformed CF-FS.

 When $ p_F > \hbar g $ we have to consider
the CF-FS as consisting of several branches belonging to several
''bands'' or Brillouin zones as is shown at Fig. 2.
The modulating potential wave  vector {\bf
g} in this case determines the size of the ''unit cell''. (If the
modulating potential is applied in one direction as in experiments \cite{12}
the ''unit cell'' is a strip extended transverse to {\bf g}). However
with this condition we may  expect some branches of the CF-FS to be
flattened. We believe this model (8) captures the important aspects of the
physics; and in addition it leads to analytical results as shown below.

\section{III. The CF conductivity tensor at $ \nu = 1/2 $}

Representing the  CFs as semiclassical particles with simultaneosly defined 
values of position and momentum we can introduce their distribution function in 
phase space $ F({\bf r,p,}t).$ For the equilibrium state of CFs the 
distribution function is the Fermi-Dirac function $ f_p$ depending on the CF 
energy $ E_p $ alone. In the presence of a small disturbance the function
$ F ({\bf r,p,} t) $ contains a small correction $ \delta n_p (r,t) $
describing its deviation from the equilibrium value. The correction can be written in the form
 $ \delta n_{\bf p} ({\bf r},t) = \displaystyle{-
(\partial f_{\bf p}/\partial E_{\bf p})g_{\bf p} ({\bf r},t)}. $ 
The dynamical response of the CFs system can then be analyzed by means of the 
Boltzmann transport equation for the non-equilibrium part of the CF 
distribution function which has the meaning of the conservation law of particle 
number in phase space.

To calculate the SAW response for the modulated 2DEG we have to solve the 
transport problem in both the local and non-local regime starting from the 
Boltzmann equation. Using linear response theory we have:
 % f10
                    \begin{equation}
\bigg(\frac{\partial}{\partial t} + {\bf v} \cdot \nabla_{\bf r} + \frac{e}{c} [{\bf v \times B(r)}] \nabla_{\bf p} \bigg) g_{\bf p} ({\bf r},t) - e {\bf E}({\bf r}, t){\bf v} = I \bf [g].
          \end{equation}
Here $ I[g] $ is the collision integral describing scattering with a
relaxation time $ \tau. $ The field $ {\bf E (r}, t) $ is the effective
electric field including both external (connected with the SAW) and
internally induced contributions.

The effect of the density modulation is taken into account through
the CF-FS deformation and through the inhomogeneity of the magnetic
field $ B({\bf r}) \ \big[B({\bf r}) = B_{eff} + \Delta B({\bf r})
\equiv B_{eff} - 4 \pi \hbar c \Delta n ({\bf r})/e \big]. $
A systematic analysis of the problem for longwave external
disturbances was performed in Refs. \cite{14} and \cite{15}. However in
the opposite case when $ ql > 1$ the problem is too complicated to be
solved straightforwardly, therefore we will proceed as follows. As a
first step we will omit the correction $ \Delta B({\bf r}) $ in order
to consider the SAW response for the CF system, whose FS is deformed
due to the density modulation, in a uniform magnetic field $ B_{eff}.$
Afterwards we will analyze semiquantitatively the influence of the
weak inhomogenious magnetic field $ \Delta B({\bf r}) $ on the
response functions and we will modify our results to include the
corresponding corrections.

We assume that ${\bf E}({\bf r},t) = {\bf E}\exp (i {\bf q r} - i \omega
t). $ Exactly at $ \nu = 1/2 $ the $ q $ and $ \omega $ dependent CF conductivity
tensor is given by
the Fourier Transform of the velocity-velocity correlator;
   %f 11
          \begin{equation}
\sigma_{\alpha \beta}^{CF} \left (\nu = \frac{1}{2} \right ) =
\frac{e^2}{(2 \pi \hbar)^2} \int
\frac{v_\alpha ({\bf p}) v_\beta ({\bf p})}{i q v_x - i \omega + 1/\tau} \frac{d\lambda}{|v|},
                      \end{equation}
where $ d \lambda = \sqrt {d p_x^2 + d p_y^2}. $ The integration has to be
performed over the CF-FS. When the CF-FS consists of several branches the
summation over the branches has to be done in Eq. (11).

At first we shall consider the contribution to the conductivity from the  flattened part  of the CF-FS $ (\sigma_{(1)}^{CF}) $   which corresponds to the model (8). We can parametrize the dispersion equation of the model as follows:
% f 12
       \begin{equation}
 p_x = \pm p_0 |\cos u|^{2/\gamma}, \qquad 
p_y = p_0 \sqrt{m_2/m_1}  \sin u ,
         \end{equation} 
where the parameter $ u $ takes values belonging to the interval
$ 0 \leq u < 2 \pi; $ and the ''+'' and ''--'' signs  are chosen
corresponding to normal domains of positive and negative values of cosine.

Taking into account the FS  symmetry we can transform the expression (11) for  $ \sigma_{(1)yy}^{CF} \,  (\nu = 1/2) $ to the form:
 % f 13
 \bea  &&
\sigma_{(1)yy}^{CF} \left (\nu  =\frac{1}{2} \right )=
\frac{8 e^2}{(2 \pi \hbar)^2}
p_0 l \sqrt{\frac{m_1}{m_2}} \frac{2 \mu^2}{(\mu + 1)^2}
 \nn \\ \nn \\ &\times&
\int \limits_0^{\pi/2}
\frac{\sin^2 u (\cos u)^{(\mu-1)/(\mu+1)}}
{1 + (q l)^2 \cos u^{4/(\mu+1)}}\, du . 
  \eea
 Here $ \mu = 1/(\gamma - 1) $ -- which defines a new
dimensionless parameter $ (0 < \mu \leq 1) .$ The value $ \mu = 1 $ corresponds to the case $ \gamma = 2 $
(the FS shaped as an ellipse). The CF mean free path $ l $ equals:
    % f 14
\begin{equation}
    l = \frac{\mu + 1}{2 \mu} \frac{p_0}{m_1} \tau.
\end{equation}
 For typical experimental  conditions \cite{6,12} $ \omega \tau << 1$ therefore $ \omega \tau $ is neglected in (13).
Making the change of variables:
% f 15
 \begin{equation}
  y = (ql)^2 (\cos u)^{4/(\mu+1)},
       \end{equation}
we  obtain the following asymptotic expression:
% f 16
\begin{equation}
 \sigma_{(1)yy}^{CF} \left (\nu = \frac{1}{2} \right ) =
\frac{4 e^2 \mu^2}{(\mu + 1)(2 \pi \hbar)^2}
\frac{p_0 l}{(q l)^\mu} \int \limits_0^{(ql)^2}
\frac{y^{\mu/2 - 1}}{1 + y} dy .
                              \end{equation}
Replacing the upper limit in the integrand (16) by infinity we have:
% f 17
\begin{equation}
 \sigma_{(1)yy}^{CF} \left (\nu = \frac{1}{2} \right ) =
b \frac{e^2}{4 \pi \hbar^2} p_0 \frac{l}{(q l)^\mu} \, .
   \end{equation}
 Here $ b = \sqrt{m_1/m_2}\, \big [4 \mu^2/(\mu + 1)\big ]
/\sin(\pi \mu/2)  $ is a dimensionless constant of the order
of unity. When the FS is a circle $ (m_1 = m_2, \; p_0 = p_F, 
\ \mu = 1) \ b = 2 $  and our result (17) coincides with the corresponding HLR (Refs. \cite{7} and \cite{8}) formula for zero flattening:
% f 18
 \begin{equation}
  \sigma_{(1)yy}^{CF} \left (\nu = \frac{1}{2} \right ) =
 \frac{e^2 p_F}{2 \pi \hbar^2 q} \, .
       \end{equation}

Comparing Eqs. (17) and (18) one can conclude that the flattening of the effective segments of the CF-FS critically changes both the magnitude and wave vector dependence of the conductivity. When the flattening is very
pronounced $ (\gamma >> 1 ) $ the parameter $ \mu $ takes values close to zero. In this case the conductivity is larger in magnitude than in the case when the CF-FS is undeformed and nearly independent of $ q.$ This result is in agreement with the experimental data \cite{12}. Thus we can  conclude that the deformed CF-FS model is supported by experiments.

When $  2p_F < \hbar g $ the entire CF-FS is described by the model (8) and Eq. (17) gives the approximation for the CF conductivity component $ \sigma_{yy}^{CF} (\nu = 1/2) $ at $ ql >> 1. $ When $ 2p_F > \hbar g $ as in experiment \cite{12} we have to  evaluate the contributions to the conductivity from the CF-FS branches (presumably nonflattened) belonging
to other ''Brillouin zones''. We shall consider them to be constructed from the pieces of the undeformed circular CF-FS. Some of them have no effective parts where $ {\bf q} \cdot {\bf v} \approx 0. $ Contributions from these CF-FS parts can be evaluated as follows:
% f 19
 \begin{equation}
 \sigma_{(2)yy}^{CF} \left (\nu = \frac{1}{2} \right ) \approx
\frac{2 e^2}{(2 \pi \hbar)^2} p_F l \sum \limits_i
\int \limits_{u_{i 0}}^{u_{i 1}} \frac{\sin^2 u du}{1 +(ql)^2 \cos^2 u}\, .
      \end{equation}

The summation in Eq. (19) has to be done over all branches of the CF-FS which have no effective parts. The limits in the integrals over $u$ are determined by range of possible values of the CF velocity component $ v_{x} $ on these branches. The upper limits $ (u_{i1}) $ take values significantly less than  $ \pi/2 $ therefore we can use the approximation $ 1 + (ql)^2 \cos^2 u \approx (ql)^2 \cos^2 u.$ It gives us:
 % f 20
 \begin{equation}
 \sigma_{(2)yy}^{CF} \left (\nu = \frac{1}{2} \right ) \approx
\frac{e^2}{ \pi \hbar^2} \frac{p_F}{q} \frac{W}{\pi q l}\, ,
      \end{equation}
 where 
  $$ \displaystyle{W = \sum \limits_i \int \limits_{u_{i0}}^{u_{i1}}} \tan^2 u du 
  $$ 
  is a dimensionless constant of the order of unity.

Those parts  of the CF-FS which have nonflattened effective parts
contribute to the conductivity as follows (summation is done over these
parts of the CF-FS):
% f 21
\begin{equation}
  \sigma_{(3)yy}^{CF} \left (\nu = \frac{1}{2} \right ) \approx
\frac{4 e^2}{(2 \pi \hbar)^2} p_F l \sum \limits_k
\int \limits_{u_{k0}}^{\pi/2} \frac{\sin^2 u du}{1 + (ql)^2 \cos^2 u}\, .
        \end{equation}
 Evaluating the integrals we have:
  % f22
 \begin{equation}
 \sigma_{(3)yy}^{CF} \left (\nu = \frac{1}{2} \right ) \approx
\frac{U e^2}{2 \pi \hbar^2} \frac{p_F}{q}
 \end{equation}
 where $ U $ is a constant of the order of unity. Comparing $
\sigma_{3yy}^{CF} (\nu = 1/2) $ and $ \sigma_{1yy}^{CF} (\nu = 1/2) $ we
can see that even in the case when the CF-FS has additional effective parts
besides the flattened segments  their contribution to the CF
conductivity for large $ q $ is smaller in magnitude than the
contribution from the flattened segments. The ratio of the magnitudes is
of the order of $ (ql)^{\mu-1}. $ Thus the contribution from the
effective segments with anomalously small curvature is the principal
term of the CF conductivity in the limit of large $ q $ for strong CF-FS
flattening $ (\gamma >> 1, \; \mu << 1).  $

The contribution from the flattened part of the CF-FS  determines
the main approximation of the CF conductivity and, consequently, the
conductivity of the 2DEG in the modulated system at $ ql >> 1 $ when
$ {\bf q} \perp {\bf g}. $ The
component $ \sigma_{xx} $ of the electron conductivity tensor which appears in the expressions  (1), (2) for the SAW velocity shift and attenuation, equals
% f 23
\begin{equation}
 \sigma_{xx} \left (\nu = \frac{1}{2} \right ) =
\frac{\rho_{yy}}{\rho_{xy^2}} \approx \frac{e^2 (ql)^\mu}
{4 b \pi p_0 l}\, .
  \end{equation}
 It follows from Eq. (23) that the 2DEG conductivity in a modulated quantum Hall system at $ \nu = 1/2 $ can be smaller in magnitude than in unmodulated systems and is almost independent of $ q $ when $ ql >> 1 $ and
flattening of the CF-FS is strong $ (\mu << 1). $

When $ {\bf g} || {\bf q} $ (the modulation applied in the ''x''
direction) the flattened segments are not effective parts of the CF-FS
(See Figs. 1 and 2). Then the main approximation to the CF conductivity component
$ \sigma_{yy}^{CF} (\nu = 1/2) $ is determined by the contribution from
the nonflattened effective parts and described by the Eq. (22).
Correspondingly the electron conductivity $ \sigma_{xx} $ is
%  f 24
 \begin{equation}
 \sigma_{xx} \left( \nu = \frac{1}{2} \right) =
\frac{e^2 q}{8 \pi U p_F}\,  .
         \end{equation}
 Under the condition $ 2 p_F < \hbar g $ (the CF-FS wholly is in the first Brillouin zone) $ U = 1 $ and the result coincides with that for the unmodulated system. Hence the effect of the density modulation on the 2DEG response is very anisotropic.

\section{IV. Magnetic field dependence of the conductivity   and the SAW anomaly near $ \nu = 1/2 $}

To analyze the CF conductivity dependence on the magnetic field $
B_{eff} $ at $\nu $ near but not equal 1/2 we
can start
from the Boltzmann transport equation for the CF distribution function
in a uniform magnetic field $ {\bf B}_{eff}. $ Following the standard
methods \cite{19} we obtain for the conductivity tensor component:
% f 25
                        \bea 
 \sigma_{\alpha \beta}^{CF} &=&
\frac{e^2 m_c}{(2 \pi \hbar)^2} \frac{1}{\Omega}
\int \limits_0^{2 \pi} d \psi \exp \left [
- \frac{i q}{\Omega} \int \limits_0^\psi v_x(\psi'') d \psi'' \right]
v_\alpha (\psi)    \nn \\ \nn \\
 &\times &
\int \limits_{-\infty}^{\psi} v_\beta(\psi') \exp \left [
\frac{i q}{\Omega} \int \limits_0^\psi v_x(\psi'') d \psi'' +
\frac{\psi' - \psi}{\Omega \tau} \right] d  \psi'. \nn\\
                              \eea
Here $ m_c, \Omega $ are  the cyclotron mass and the cyclotron frequency
at the field $ B_{eff}, \psi $ is the angular coordinate of the CF
cyclotron orbit $ (\psi = \Omega t, \; t $ is the time of the CF motion
along the cyclotron orbit). We have taken $ \omega \tau << 1. $ 

For  weak magnetic fields $ (\Omega \tau << ql) $ we can evaluate the 
integrals over $ \psi $ and $\psi' $ in Eq. (25) by means of the method of 
stationary phase. As a result we obtain the lowest order terms of the 
expansion of the conductivity in powers of the small parameter $ \Omega \tau / 
q l. $ For undeformed (circular) CF-FS the main terms in the expansions give 
the correct expressions for the conductivity components in the limit of zero 
magnetic field. However when the CF-FS is noncircular and has flattened parts 
the main approximation to the conductivity component $ \sigma_{yy}^{CF} $ 
diverges in the limit of zero field. To evaluate the CF conductivity so
that we can pass smoothly to the $ B_{eff} \to 0$ limit for a flattened
CF-FS we proceed \cite{23} as follows. Express the velocity components$ v_\beta(\psi') $ as Fourier series: 
   % f 26
                                 \begin{equation}
v_\beta (\psi') = \sum \limits_k v_{k \beta} \exp(i k \psi').
                                 \end{equation}
Substituting Eq. (26) into Eq. (25) we obtain:
 % f 27
                \bea
\sigma_{\alpha \beta}^{CF}& =& \frac{e^2 m_c}{(2 \pi \hbar)^2}
\sum \limits_k v_{k \beta} \int \limits_0^{2\pi}
d \psi v_\alpha(\psi) \exp(i k \psi)
    \nn \\   
 &    \times &
\int \limits_{-\infty}^0 \exp \bigg [(i k \Omega + i q v_x(\psi)+ 1/\tau ) \theta 
  \nn \\
&+ & i q \int \limits_0^\theta  (v_x(\psi + \Omega \theta')
- v_x(\psi)) d \theta' \bigg ] d \theta .
         \eea 
 Here $ \theta = (\psi' - \psi)/\Omega. $

Introducing a new variable $ \eta, $ which is defined by the expression:
 % f 28
  \bea
\eta &\equiv& \left (- i \omega + 1/\tau + i k \Omega +
i q v_x (\psi) \right ) \theta 
\nn \\ 
 &+& i q \int \limits_0^\theta
\big [ v_x(\psi + \Omega \theta') - v_x (\psi) \big ] d \theta',
                   \eea
 we arrive at the result
   % f 29
  \bea
\sigma_{\alpha \beta}^{CF}& = & \frac{i e^2}{(2 \pi \hbar)^2} m_c
\sum \limits_k v_{k \beta}
\int \limits_{-\infty}^0 e^\eta d \eta 
 \nn \\ & \times & 
\int \limits_0^{2 \pi} \frac{v_\alpha (\psi) \exp(i k \psi)}{\omega + i/\tau - k \Omega -
q v_x(\psi + \Omega \theta)} d \psi.
  \eea
 When the CF-FS consists of several branches the summation over the branches has to be done in Eq. (29). 

\section{a. Case $\bf g \perp q $}

It was shown in the previous section that for ${\bf q} \perp {\bf g} $ the main contribution to the conductivity
arises from the flattened parts of the CF-FS. To evaluate this contribution we transform the integration over $ \psi $ to an integration over the CF-FS:
% f 30
   \begin{equation}
   m_c \int \limits_0^{2\pi} d \psi = \int \frac{d \lambda}{|v|} \, .
                              \end{equation}
Under the conditions $ ql >> 1, \; \omega \tau << 1, \; \Omega \tau < 1$ the  variable $ \theta $ is approximately equal to
$ \eta \tau (1 + i q v_x \tau)^{-1}. $ Expanding the last term in the denominator (29) in powers of $ \Omega \theta $ and keeping the first terms in the expansion one  obtains
  \bea % f31
 q v_x (\psi + \Omega \theta) &=&
q v_x (\psi) + \eta \Omega \tau q(d v_x/d \psi)(1 + i q v_x \tau)^{-1}
   \nn \\    \nn \\             
&+& \frac{\eta^2}{2}(\Omega \tau)^2 q (d^2 v_x/ d\psi^2)(1 + i q v_x \tau)^{-2}
  \nn \\ \nn \\ &  \equiv &
   q v_x(\psi) - \eta \delta_1 (q,\psi) +
i \eta^2 \delta_2 (q,\psi).
                              \eea
We can estimate the averages of the functions $ \delta_1 (q, \psi) $ and $ \delta_2 (q, \psi) $ over the CF-FS. These are:
   % f 32
   \bea
<\delta_1 (q,\psi)> &=& \frac{1}{2 \pi} \int\limits_0^{2 \pi}
\delta_1 (q,\psi) d \psi = 0;
   \nn \\ 
<\delta_2 (q,\psi)>  &= &\frac{1}{2 \pi} \int\limits_0^{2 \pi}
\delta_2 (q,\psi) d \psi \approx \frac{(\Omega \tau)^2
\overline \delta (q)}{\tau } \, .
                              \eea
  This defines $ \overline\delta(q) $. 

Using the model (8) we can prove that for 
moderately flattened 
CF-FS the function $ \overline \delta (q) $ takes positive values of the
order of $ (ql)^{-(1 + \mu)/2} $ (for underformed CF-FS it is of the
order of $ (ql)^{-1} ).$ For strong flattening of the CF-FS
$ \overline \delta (q) $ is practically independent of $ q $ and takes
negative values of the order of unity.
Using this estimate we obtain the approximation:
 % f 33
                    \begin{equation}
 \sigma_{1yy}^{CF} = \frac{1}{2} \sigma_{1yy}^{CF}
\left (\nu = \frac{1}{2} \right)
\big [S_\mu^+(\Omega \tau) + S_\mu^-(\Omega \tau) \big].
 \end{equation}
Here $ \sigma_{1yy}^{CF} (\nu = 1/2) $ is the conductivity at one half
filling (17) and the functions $ S_\pm (\Omega \tau) $ are:
 % f 34
                    \begin{equation}
 S_\mu^\pm(\Omega \tau) = \int \limits_{- \infty}^0 e^\eta
 \big[ \kappa_\pm (\eta) \big ]^{\mu - 1} d \eta ,
        \end{equation}
 where $ \kappa_\pm (\eta) = 1 \mp i \Omega \tau
(1 \mp i \Omega \tau \overline \delta \eta^2).$

The approximation (33), (34) represents the main term in the
expansion of  the conductivity in inverse powers of the large
parameter $ q l. $ When $ \mu < 1 $ the next term of the expansion equals:
% f 35
                    \begin{equation}
\frac{1}{2} \sigma_{1yy}^{CF} \left (\nu = \frac{1}{2} \right )
\frac{i}{q l} \tan \frac{\pi \mu}{2}
\Big [ S_{\mu+1}^-(\Omega \tau) - S_{\mu+1}^+(\Omega \tau) \Big ].
 \end{equation}
 For $ \mu = 1 $ the first correction is of the order of $\ln(ql)/ql. $

For small $\overline \delta \Omega \tau $ one can expand the function $ S_\mu^\pm (\Omega \tau) $ in powers of $ \Omega \tau .$ Keeping  terms smaller than $(\Omega\tau)^3$ one has
%         f 36
 \begin{equation}
{\sigma}_{1yy}^{CF} = {\sigma}_{1yy}^{CF} (\nu = 1/2)
\left [1-a^{2}(\Omega\tau)^2 +  \xi (\Omega\tau)^2 \right] ,
                          \end{equation}
 where $ a^{2}=(1-\mu)(2-\mu)/2 $ and $ \xi = 4(1-\mu)\overline{\delta}. $

For  moderate flattening of the effective parts of the CF-FS the
constant $ \xi $ (positive in this case) is small compared to $ a^2 $
because of the small factor $ \overline \delta.$ For  strong
flattening of the CF-FS this constant $\xi$ takes negative values of the
order of $ a^2.$ In both cases we can omit the last term in 
Eq. (36)
neglecting it for $ \mu \sim 1, $ or combining it with the previous term
(for $ \mu << 1 $).
The contributions from the other (nonflattened) parts of the CF-FS to
the CF conductivity component $ \sigma_{yy}^{CF} $ are small compared to
$ \sigma_{1yy}^{CF} $ as  follows from  Eqs. (20), (22). Hence Eq.
(36) gives the main approximation for $ \sigma_{yy}^{CF} $ in modulated
system at $ \nu $ near  1/2.
Other components of the CF conductivity tensor can be evaluated 
similarly. For $ ql >> 1$ we obtain  $
\sigma_{xx}^{CF} \sim \sigma_0 /(ql)^2; \; \sigma_{yx}^{CF}  = -
\sigma_{xy}^{CF} 
\sim \Omega \tau \sigma_0/(ql)^{1 + \mu}$ where $\sigma_0 =
\sigma_{1yy}^{CF}(\nu = 1/2)(ql)^\mu = \displaystyle{ b
(e^2/4\pi \hbar^2) p_0 l.}$ Hence under conditions of the
experiment [12] $ (\Omega \tau < 1) \; (\sigma_{xy}^{CF})^2 /
\sigma_{xx}^{CF}  \sigma_{yy}^{CF}
\sim (\Omega \tau)^2/(ql)^\mu << 1 $ and we
can calculate the electron conductivity using the formula:
    % f 37 
               \bea
\sigma_{xx}(q) & =& 
\frac{e^4}{(4 \pi \hbar)^2} \rho_{yy}^{CF}
= \frac{e^4}{(4 \pi \hbar)^2}
\frac{\sigma_{xx}^{CF}}{\sigma_{xx}^{CF}\sigma_{yy}^{CF} +
(\sigma_{xy}^{CF})^2}
  \nn \\ & \approx &
\frac{e^4}{(4 \pi\hbar)^2} (\sigma_{yy}^{CF})^{-1}.
                              \eea
Substituting the result into the formulas (1) and (2) we obtain the following approximation $ (\omega \tau << 1, \ ql >> 1, \ \Omega \tau < 1): $
   % f38
                         \begin{equation}
 \frac{\Delta s}{s} = \frac{\alpha^2}{2}
\frac{1}{1 +  \tilde \sigma^2}
\left( 1 - \frac{ 2 a^2 \tilde \sigma^2}{1 + \tilde \sigma^2} (\Omega
\tau^2) \right);
                         \end{equation}
% f 39
                         \begin{equation}
\Gamma = q \frac{\alpha^2}{2} \frac{\tilde \sigma}{1 + \tilde \sigma^2}
\left( 1 - \frac{ a^2 \tilde \sigma^2}{1 + \tilde \sigma^2} (\Omega
\tau^2) \right).
                         \end{equation}
Here 
  $$
  \displaystyle{ \tilde \sigma = \frac{\sigma_{xx} (\nu =
1/2)}{\sigma_m} \equiv \left( \frac{e^2}{4b\pi} \frac{(ql)^\mu}{p_0l} \right ) \frac{1}{\sigma_m}. }
  $$

For a strong flattening of the CF-FS $ (\mu << 1) $ we have to
replace $ a^2 $ by $ a^2 + |\xi|.$

\section{b. Comparison with Experiments of Ref. \cite{12}}

\begin{figure}[t]
\begin{center}
\includegraphics[width=5.cm,height=7.cm]{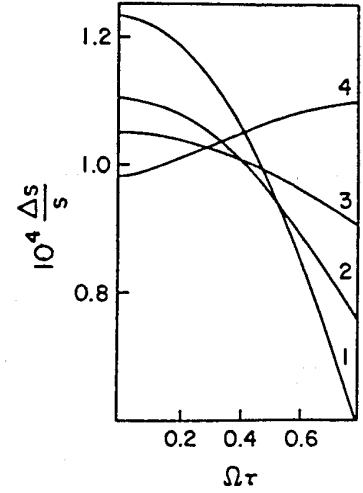}
\caption{The SAW velocity shift $\Delta s/s$ as a function of $\Omega\tau $ calculated for the following values of the parameters: $ n = 0.7 \cdot 10^{11} $cm$^{-2}; \ \tau = 2 \cdot 10^{-11}$ s; $q = 9 \cdot 10^{4}$ cm$^{-1}; \ \alpha^2/2 =
3.2 \cdot 10^{-4}; \sigma_m = 0.6 \cdot 10^6$ cm/s \cite{25}; $ \mu = 3/4 $ (1), 1/2 (2), 1/4 (3). The curve (4) represents $ \Delta s /s $ as a function of $\Omega \tau$ for the circular (undeformed) CF-FS. }
\label{rateI}
\end{center}
\end{figure}

Expressions (38) and (39) are the new results in our theory. They predict peaks both in the SAW attenuation
and velocity shift at one half filing. This follows from the fact that
$ \Omega = 0 $ at $ \nu = 1/2 $ when $ B_{eff} = 0. $ The peaks arise
due to the distortion of the CF-FS in the presence
of the electron density modulation. These peaks in $\Delta s/s$
for several values of the parameter  $\mu $ are shown in Fig. 3. For
$ \alpha^2/2 = 3,2 \cdot 10^{-4}$ in GaAs, $ n = 0,7 \cdot 10^{11}$
cm$^{-2}, \ \tau \sim 10^{-11}$ s, $ q \sim 10^5$ cm$^{-1},$ which
agree with the conditions of the experiments \cite{12}, amplitudes of the
peaks can reach the order of $10^{-4}$ when the parameter $ \mu$
characterizing the rate of the distortion of the CF-FS is small enough.
This order of  magnitude of the peaks in $ \Delta s/s$ agrees with
 the experimental data of Ref. \cite{12}.
When the CF-FS
flattening is strong $ (\mu << 1) $ the magnitude of the peak of the
velocity shift is practically independent of the SAW wave vector $ q.$
Also these anomalies are not sensitive to any relation between $ q$
and the density modulation wave vector $ g.$ The effect is strongly
anisotropic. The peaks can arise only in the geometry when
$ {\bf g}\perp {\bf q} $ and the CF-FS effective parts coincide with
its segments flattened due to the density modulation.

\section{c. Case $\bf q || g$}

When $ {\bf g} || {\bf q} $ the effective parts of the CF-FS are
nonflattened $ (\mu = 1). $ Then main term in the expansion of the CF conductivity in inverse powers of $ ql $ is independent of the magnetic field. When we take into account the next term of the expansion we arrive at the following expression after a lengthy, but straightforward calculation:
 % f B16- 40
                         \begin{equation}
\sigma_{yy}^{CF} = \frac{2 e^2 p_F^2 \tau}{(2 \pi \hbar)^2 m^*}
\int \limits_{-\infty}^0 e^\eta d \eta
(I_1^+ + I_1^- + I_2^+ + I_2^-)
   \end{equation}
 where
 % f B17      41-42
                         \bea &&
I_1^\pm = \kappa_\pm \int \limits_0^{\pi/2}
\frac{\sin^2 \psi d \psi}{(\kappa_\pm)^2 + (q l)^2 \cos^2 \psi}\, ,
   \\ \nn \\ &&
I_2^\pm  = \pm i \kappa_\pm \int \limits_0^{\pi/2}
\frac{\sin \psi \cos \psi d \psi}{(\kappa_\pm)^2 + (q l)^2 \cos^2 \psi}\, .
       \eea
 In the limit $ \Omega \tau << 1 $ the sum $ I_2^+ + I_2^- $ becomes zero and we have:
  % f B20           43
                         \bea
\sigma_{yy}^{CF}& =& \frac{2 e^2 p_F^2 \tau}{(2 \pi \hbar)^2 m^*} \int \limits_{-\infty}^0 e^\eta \bigg [\frac{\pi}{q l} 
   + \frac{2 \ln (q l)}{(q l)^2} \Omega \tau
\nn \\ \nn \\ & + & (\Omega \tau)^2 f (\eta) O
\left (\frac{1}{q l} \right) \bigg ] d \eta.
          \eea
 where $ O $ means ''order of''.

Now, noting that the first two terms are independent of $ \eta, $ and retaining only them in the integral, we obtain the CF conductivity tensor component:
 % f 44
                         \begin{equation}
 \sigma_{yy}^{CF} = \sigma_{yy}^{CF}  \left( \nu = \frac{1}{2} \right) \left[1 + \Omega \tau \frac{2\ln (ql)}{\pi ql} \right],
                         \end{equation}
 where $ \sigma_{yy}^{CF} (\nu = 1/2) $ is described by Eq. (17). Expression (42) describes the CF conductivity increasing as $B_{eff} $ increases. This corresponds to the minimum in the SAW velocity shift at $ \nu = 1/2. $ This minimum was observed repeatedly in non-modulated FQHE systems \cite{6}.

\section{V.  Effect of the modulation of the magnetic field }

To analyze the effect of the inhomogeneity of the effective magnetic field arising in modulated CF systems due to the density modulation we shall start from the Lorentz force equations describing the CF motion along the cyclotron orbit:
  % f 45
                         \begin{equation}
\frac{dp_x}{dt} = - \frac{|e|}{c} B({\bf r}) v_y, \qquad
\frac{dp_y}{dt} =  \frac{|e|}{c} B({\bf r}) v_x , 
                         \end{equation}
where $ B({\bf r})= B_{eff} + \Delta B({\bf r}) \equiv B_{eff}
- 4 \pi \hbar c
\Delta n ({\bf r}) / e. $ For a single-harmonic density modulation along
the ''y'' direction we can write $ \Delta n ({\bf r}) \equiv \Delta n(y)=
\Delta n \cos (gy). $  In the  following the correction term
$ \Delta B(\bf r)$ is assumed to be small compared to $ B_{eff}. $
Under this condition we can write the CF velocity {\bf v} in the form
$ {\bf v}= {\bf v_0} + \delta{\bf v}, $ where ${\bf v}_0 $ is the
uniform-field velocity and $ \delta {\bf v} $ is a small correction
term arising in the presence of the inhomogeneous magnetic field
$\Delta B({\bf r}). $

At first we will suppose that the CF-FS is undeformed (a circle). In this case we have: $ v_{0x} = v_F \cos \Omega t; \  v_{0y} = v_F \sin \Omega t; \
\Delta n (y) \approx \Delta n \cos (gY - g R \cos \Omega t), $
where $ v_F$ is the CF's Fermi velocity, $ R = v_F/\Omega $ is the radius of the cyclotron orbit and $ Y $ is the  ''y'' coordinate of the guiding center. Substituting the expressions for {\bf v} and $ \Delta n(y) $ into Eqs. (45) and keeping only first-order terms we obtain:
 % f 46
           \bea                
\frac{d(\delta v_x)}{dt} & =& - \Omega \delta v_y - \frac{\Delta B}{B_{eff}} v_F
\sin \Omega t \cos \big(gY - g R \cos (\Omega t)\big), 
                         \nn \\ \nn \\
\frac{d(\delta v_y)}{dt}& =&   \Omega \delta v_x + \frac{\Delta B}{B_{eff}} v_F
\cos \Omega t \cos \big(gY - g R \cos (\Omega t)\big) .\nn \\
                         \eea
   It is natural to assume that to the first order in the modulating field the corrections $ \delta v_x $ and $ \delta v_y $ are periodic over the unperturbed cyclotron orbit. This assumption is equivalent to that used in Ref. \cite{17}. See {\it Note added} (before References).
 Using these equations we calculate the averages of the corrections $ \delta v_x $ and $ \delta v_y $ over the cyclotron orbit and obtain the expressions for
the components of the velocity of the guiding center $ V_x $ and $
V_y. $ defined below. Expanding the functions $ \cos (g R \cos \psi) $ and
$ \sin (g R \cos \psi) $ in Bessel functions and substitung these
expansions into Eqs.(46) we arrive at formulas 
  % f 47
        \bea
 V_x & \equiv & \big <\delta v_x \big > = - \frac{v_F}{2 \pi}
\frac{\Delta B}{B_{eff}} 
\int \limits_0^\psi \cos \psi \cos (g Y - gR \cos \psi) d \psi
 \nn \\ \nn \\ & = &
  - v_F \frac{\Delta B}{B} \sin g Y J_l (gR); 
    \nn \\ \nn \\
  V_y & \equiv &  \big <\delta v_y \big >
 = - \frac{v_F}{2 \pi} \frac{\Delta B}{B_{eff}} 
\int\limits_0^\psi \sin \psi\cos (g Y - gR \cos \psi)d \psi
\nn \\ & = &0.
                         \eea
Here $ \psi = \Omega t .$

The formula (47) describe the drift of the guiding center along the direction perpendicular both to the magnetic field and to the electric field producing the density modulation in the 2DEG. Similar effects in the periodically modulated 2DEG in low magnetic fields were considered before \cite{17,18}.

To evaluate the contribution from the guiding center's drift to the CF conductivity semiquntitatively for the extended range of values of the perturbing wave vector $ q $ including the nonlocal region $ (ql > 1), $ we will make the following assumption. We assume that in the presence of a weak inhomogeneous magnetic field the components of the CF velocity can be writen $ v_x = v_{x_0} + V_x, \   v_y = v_{y_0} + V_y.$ Using these expressions, we can evaluate the CF conductivity as follows:
 % f48
 \begin{equation}
 \sigma_{\alpha \beta}^{CF} \approx \frac{g}{2\pi}
\int \limits_{-\pi/g}^{\pi/g}
\sigma_{\alpha \beta}^{CF} (Y) d Y
         \end{equation}
 where
   \bea %f49
\sigma_{\alpha \beta}^{CF} (Y) &=& \frac{i e^2 m_c}{(2 \pi \hbar)^2} \sum \limits_k (v_{k \beta 0} + V_\beta (Y) \delta_{k 0}) \int \limits_{- \infty}^0 e^\eta d \eta
   \nn \\ \nn \\ & \times & \!\!
 \int \limits_0^{2 \pi} \!\!
\frac{[v_{\alpha 0}(\psi) + V_\alpha (Y)] \exp (i k \psi)}
{\omega + i/\tau - k \Omega - q v_{x 0} (\psi + \Omega \theta)
- q V_x (Y)} d \psi. \nn \\
         \eea

The correction $ V_x$ produces additional terms in the expressions for the $ \sigma_{\alpha \beta}^{CF}. $ When the CF-FS is a circle a nonzero correction arises only in the expression for $\sigma_{xx}^{CF} .$ For  small $ q  \; (ql << 1) $ we have:
  % f 50
           \bea
  \sigma_{xx}^{CF} &=& \frac{1}{2} \sigma_0
\left (1 + 2 \frac{\delta \sigma}{\sigma_0} \right) 
   \nn \\ \nn \\ &\equiv&
\frac{1}{2} \sigma_0 \left \{\big[1 + (\Omega\tau)^2\big]^{-1} + 2
\left (\frac{\Delta B}{B_{eff}}\right)^2 
J_1^2(gR) \right \} . \nn \\
         \eea 
 The remaining components of the CF conductivity tensor are the same as in the absence of the density modulation and equal correspondingly $ {
\sigma_{yy}^{CF} = \frac{1}{2} \sigma_0 \big[1 + (\Omega \tau)^2 \big]^{-1}, \
\sigma_{xy}^{CF} = -\sigma_{yx}^{CF} =\sigma_{yy}^{CF} \Omega \tau}.$  These results are valid for arbitrary values of the parameter $ \Omega \tau. $

It should be noted that the expression (50) may be obtained without using Eqs.(48), (49). Following Ref. \cite{24} we can consider the guiding center drift
to lead to a diffusion along the ''x'' direction with diffusion
coefficient $\delta D $ which equals:
% f 51
           \begin{equation}
\delta D = \tau \frac{g}{2\pi} \int \limits_{-\pi/g}^{\pi/g}
V_x^2 (Y) dY.
         \end{equation}
The term $ \delta D $ is the additional contribution to the $ xx $
component of the diffusion tensor. The latter is connected with the
conductivity through the Einstein relation $\sigma_{\alpha\beta} = Ne^2 D_{\alpha\beta} \; (N = m^*/(2\pi \hbar^2) $ is the CF density of states). Substituting Eq.(51) into this relation we obtain for $\delta \sigma $ the expression, which coincides with the second term in Eq.(51). This coincidence confirms that we may use the formulae (48), (49) for the evaluation of the order of magnitude of the corrections
arising in the conductivity due to the inhomogeneity of the magnetic field.

The electron conductivity $ \sigma_{xx} $ can be evaluated as 
\\\\
$ \displaystyle{ \left( \frac{\Delta B}{B_{eff}} \Omega \tau =
\frac{\Delta n}{n} \frac{p_F l}{\hbar} = \frac{\Delta n}{n}k_F
l \right) :} $
% f 52-53
  \bea 
\sigma_{xx} &= &\frac{e^4}{(4\pi \hbar)^2} \rho_{yy}^{CF} =
\frac{e^4}{(4\pi \hbar)^2} \frac{2}{\sigma_0} 
  \nn \\ \nn \\ &\times& 
 \left \{
1 + 2 \left (\frac{\Delta B}{B_{eff}}\Omega \tau \right)^2
\frac{J_1^2(gR)}{1 + 2 (\Delta B/B_{eff})^2 J_1^2(gR)}\right \} 
   \nn \\ \nn \\ 
&\approx& \frac{e^4}{(4\pi \hbar)^2}  \frac{2}{\sigma_0}
\left[ 1 + 2 \left(\frac{\Delta n}{n}k_F l \right )^2 J_1^2 (gR) \right ],
                      \\  \nn \\
\sigma_{yy} &  =& \frac{e^4}{(4\pi \hbar)^2} \rho_{xx}^{CF} =
\frac{e^4}{(4\pi \hbar)^2} \frac{2}{\sigma_0} 
  \nn \\ \nn \\  & \times &
\left (\frac{1}{1 + 2 (\Delta B/B_{eff})^2 J_1^2(gR)} \right )
 \approx \frac{e^4}{(4\pi \hbar)^2}  \frac{2}{\sigma_0} .
                          \eea
  The correction to the conductivity component $ \sigma_{xx} $ is of the same order as the corresponding result of Ref. \cite{14}. It confirms once more that the simplified expression assumed here may be used to evaluate the CF conductivity.

When the density modulation is weak  $ \big[(\Delta n/n)
k_{F}l<<1 \big]  $ the corresponding correction to the conductivity $ \sigma_{xx} $ is small and we can neglect it. In this case the inhomogeneity of the effective magnetic field does not significantly effect dc transport. For  stronger modulation $ \big [(\Delta n/n)k_{F}l\sim 1 \big] $
the conductivity component $\sigma_{xx}$ is appreciably changed.

The magnetic field dependence of the electron conductivity is determined by the value of the parameter $ gR $. When $ gR>>1,$  the correction to the conductivity is small in magnitude [becase of the small  parameter $ (gR)^{-1} ] $ and describes   the Weiss oscillations effect for CFs. For $ gR<1 $ the
function $ J_1^2(gR) $ takes values of the order of unity and increases upon  increasing of $ B_{eff} $ (it corresponds to the decrease of the parameter $ gR $). Hence for $ gR<1 $ and $ (\Delta n/n) k_{F}l\sim 1  $ the expression (52) describes a large minimum in the magnetic field dependence of the electron conductivity $ \sigma_{xx} $ (or magnetoresistivity $ \rho_{yy} $) around $ \nu =1/2 $ similar to the corresponding result of Ref. \cite{14}. Such minimum of the longitudinal magnetoresistivity in a modulated 2DEG for a weak modulation
perpendicular to the current direction was observed recently \cite{13}. According to the conditions of these experiments $ \big [\Delta n/n)\sim 10^{-2}, \  n\sim 1,8 \cdot 10^{11} $cm$^{-2},\  l\sim 1,4 \cdot 10^{-4}$cm, $ g\sim 10^{5}$cm$^{-1} \big ] $ the effect occures when the parameters $ (\Delta n/n) k_{F}l $ and $ gR $ are of the order of unity, which agrees with the theoretical estimates. The
theoretical results obtained in Ref. \cite{14} reproduce the width and the magnitude of the grating induced minimum in the magnetoresistivity rather well. Our estimation based on the expression (52) also gives a reasonable agreement with these experimental results (See Fig. 4).

\begin{figure}[t]
\begin{center}
\includegraphics[width=5.5cm,height=7cm]{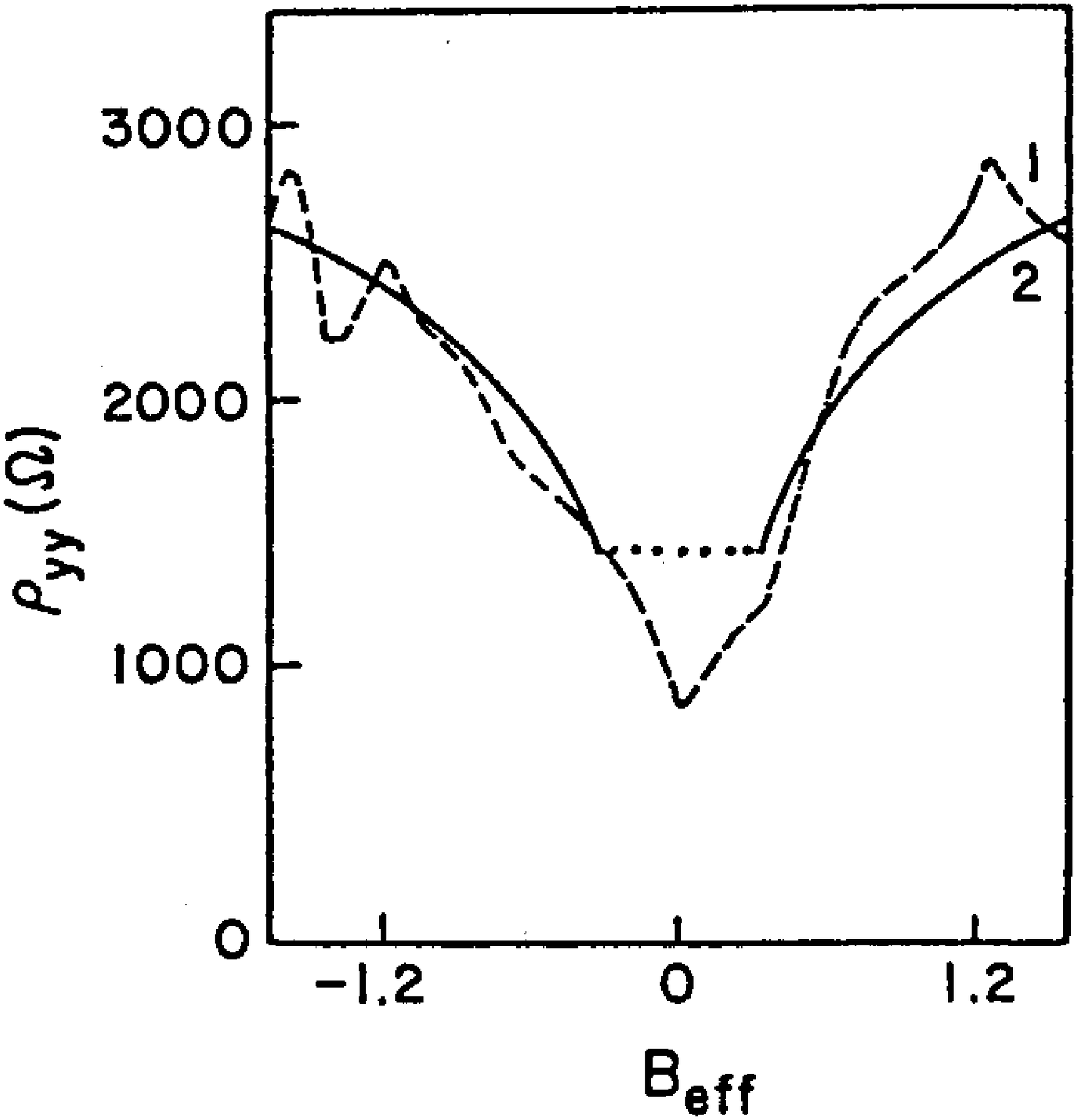}
\caption{dc magnetoresistivity as a function of the effective magnetic field.
Dashed line (1) -- experiment of Ref.[13], solid line (2) -- theory for the parameters $ n = 1.8 \cdot 10^{11} $ cm$^{-2}; \ \Delta n /n = 0.01; \ g = 10^5$ cm$^{-1}; \ l = 1.4 \cdot 10^{-4}$ cm; $\sigma_0^{-1} = 270 \Omega.$ The dotted part of the curve (2) corresponds the the region of
the values of $ B_{eff}$ where Eq.(52) can not be applied. 
 }
\label{rateI}
\end{center}
\end{figure}

Thus the analysis performed above  corroborates that the inhomogeneity of $ B_{eff} $ originating from the density modulation is the main factor determining the magnetic
field dependence of the 2DEG conductivity around one half filling for small $ q\ (ql << 1) $. This is similar to the results obtained in Ref. \cite{14}.  However this conclusion  can not be extended to the  {\it nonlocal} region of large $ q\ (ql > 1) $ as we show below.

Straightforward calculations starting from Eqs. (48) and (49) show that for the undeformed CF-FS, with large $ q\ (ql > 1) $ in the same way as for small $ q,$ only $ \sigma_{xx}^{CF} $ changes due to the inhomogeneity of the magnetic field along the ''y'' direction. However the correction to $ \sigma_{xx}^{CF} $ does not significantly influence  the CF resistivity 
$\rho_{yy}^{CF}$ (and, consequently the corresponding
 component of the electron conductivity $ \sigma_{xx} $) close to $\nu=1/2$. This follows from the expression (37) for $ \sigma_{xx} $. The ratio $ (\sigma_{xy}^{CF})^{2}/\sigma_{xx}^{CF}\sigma_{yy}^{CF} $ for $ql>>1$ is of the order of $ (\Omega\tau)^{2}/ql $. Therefore  when $ \Omega \tau < \sqrt{ql} $  we can use the approximation $ \rho_{yy}^{CF} \approx ( \sigma_{yy}^{CF})^{-1}  $ and $ \sigma_{xx} $ does not depend on $ \sigma_{xx}^{CF} $. Hence near one half filling where $\Delta \nu = |\nu - 1/2| $ is less than $\sqrt{(q/k_F)(k_F l)^{-1}} $ (in fact under the conditions of the experiment \cite{12}
$ \sqrt{(q/k_F) (k_F l)^{-1}} \sim 0,1) $ the
correction arising due to the inhomogeneity of the magnetic field does not contribute to the response of the modulated 2DEG system to the SAW propagating perpendicular to the direction of the density modulation. The width of the region around $ \nu=1/2 $ where $ \sigma_{xx} $ is independent of $ \sigma_{xx}^{CF} $ $ (\Delta\nu\sim 0,1) $ is of the same order as the width of the anomalous peak observed in Ref. \cite{12}. Thus this peak is wholly (or in  major part) in the region near $ \nu=1/2 $ where the inhomogeneity of $ B_{eff} $ due to the grating can not influence the response of the 2DEG with the undeformed (circular) CF.

At larger $ \Delta \nu $ when $ \Omega \tau \sim \sqrt{ql} $ the
influence of the density modulation on the electron conductivity
becomes significant. When $ B_{eff} $ is large enough to satisfy the condition $ (\Delta B/B_{eff}) ql = (\Delta
n/n) k_F l qR << 1, $ we obtain the following asymptotic expression for $ \sigma_{xx}: $
% f 54
                         \bea
\sigma_{xx} & =& \frac{e^4}{(4\pi \hbar)^2} \rho_{yy}^{CF}(\Delta n=0) 
  \nn \\ \nn \\ & \times & 
 \left \{
1 -  \left (\frac{\Delta n}{n} k_F l \right)^2 (qR)^2 J_1^2(gR)
 \right \}.
                          \eea
It is clear from the comparison of Eqs. (52) and (54) that the
corrections to the conductivity due to inhomogeneity of the effective magnetic field in the local $ (ql << 1) $ and nonlocal $ (ql > 1) $ limits are of  opposite signs. This difference arises due to the distinctions between contributions due to the guiding center drift under the local and nonlocal regimes.

In the region of small $ q $ the
diffusion due to the guiding center drift enhances the conductivity.
In the opposite limit $ (ql >> 1) $ the main term in the conductivity
arises from the effective parts of the FS where $ {\bf q \cdot v}
\approx 0 $ (at $ \omega \tau << 1). $ The existence of a nonzero average
of the velocity component $ v_x $ over the cyclotron orbit prevents
satisfying this condition and leads to decrease of the
conductivity. Positive corrections to the conductivity which arise
from the diffusion are smaller in magnitude than negative. The ratio
of magnitudes is of the order of $ (ql)^{-1} $ for the 
2DEG system  with the undeformed CF-FS.

It is known from SAW experiments in unmodulated 2DEG in
the quantum Hall regime \cite{6} and from theory \cite{7} [See also Eq. (44)] that $ \rho_{yy}^{CF}(\Delta n = 0)$ and, correspondingly, $ \sigma_{xx}$ has a maximum at $ \nu = 1/2 $ and decreases when $ \Delta \nu $ grows. It corresponds to the well known minimum in the SAW velocity shift. In the region where Eq. (54) can be applied $(\Omega\tau >\sqrt{ql})$
the parameter $ gR $ is of the order or less than $ \sqrt{ql} $ (in agreement with the experiments of Ref. \cite{12} we assume that $q\sim g$).  When $ qR <  1$ the function $ (gR)^{2} J_1^2(gR) $  decreases upon  increasing $B_{eff}$. Hence the trends of the  modulation independent factor $ \rho_{yy}^{CF}(\Delta n = 0)$ and the modulation-dependent factor in Eq. (54) are opposite when $ gR < 1 $. Under some
conditions it can lead to increase of the conductivity and a
corresponding decrease of the SAW velocity shift for
increase of the effective magnetic field. However it has to be
stressed again that the formula (54) is valid only for $ \Omega \tau > \sqrt{ql} .$ At $ \Omega \tau < 1$ the conductivity $ \sigma_{xx} $ coincides with that for the unmodulated system and (for undeformed CF-FS) exhibits a minimum at $ \nu = 1/2.$

The analysis performed above shows that we can not obtain a correct theoretical description of the SAW anomaly observed near $ \nu = 1/2 $ for $ ql > 1 $ neglecting the deformation of the CF-FS in modulated 2DEG. The inhomogeneous
contribution to the field $ B_{eff} $ by itself does not cause this anomaly in the SAW velocity shift. This conclusion disagrees with the results of the study of the problem of the SAW anomaly performed  in Refs. \cite{14} and \cite{15}. The results of these works \cite{14,15} are
valid when $ ql << 1 $ and can not be extended to the nonlocal regime where the CF conductivity strongly depends on the CF-FS local geometry. A minimum in the electron conductivity analysed in Refs. \cite{14} and \cite{15} can be identified with that observed in the 2DEG dc magnetoresistivity (See Ref. \cite{13}) but not with that responsible for the SAW anomaly reported
in Ref. \cite{12} because the latter was observed at large $ q $ ($ql > 1).$

To analyse semiquantitatively the effect of the inhomogeneity of the magnetic field on the 2DEG conductivity taking into account the distortion of the CF-FS due to the density modulation we will assume at first that $ (\Delta B/B_{eff}) ql << 1. $ Also we will assume here that the velocity of the guiding center drift is of the same order as in the system whose CF-FS is undeformed, because a small deformation of the effective parts of the CF-FS may not result in significant changes of the averages over the CF-FS for this particular
case.

When the CF-FS is deformed the component of the CF conductivity
$ \sigma_{yy}^{CF} $ changes in the presence of the inhomogeneus
magnetic field. Using Eqs.(49), (50) we obtain:
% f 55
                         \bea
\sigma_{(1)yy}^{CF} &\approx& \frac{1}{2} \sigma_{(1)yy}^{CF} (\nu = 1/2) \frac{g}{2 \pi} 
  \nn \\ \nn \\ &\times&
\int \limits_{- \pi/g}^{\pi /g} dY
\big[ S_\mu^+ (\Omega \tau, Y) + S_\mu^- (\Omega \tau, Y) \big ],
                          \eea
 where
% f 56
                         \begin{equation}
S_\mu^\pm (\Omega \tau, Y) = \int \limits_{-\infty}^0 e^\eta
\left (\kappa_\pm (\eta) + i q l \frac{\Delta B}{B_{eff}}
\varphi (Y) \right)^{\mu -1} d \eta
                          \end{equation}
 and
% f 57
                         \begin{equation}
\varphi(Y) =  \sin(gY) \Phi_1 (g, B_{eff}).
                          \end{equation}
 The functions $ \Phi_0 $ and $\Phi_1 $ are of the order or less
than unity. For the undeformed CF-FS these two functions coincide with Bessel functions [See Eq. (47)].

For $ \Omega \tau < 1 $ we have:
% f 58
                         \bea
\sigma_{(1)yy}^{CF} & \approx &
 \sigma_{(1)yy}^{CF} \bigg(\nu = \frac{1}{2}
\bigg) \bigg \{ 1 - a^2 \bigg[ (\Omega \tau)^2
   \nn \\ \nn \\ & + &
\frac{1}{2}\left(ql \frac{\Delta B}{B_{eff}} \right)^2 \Phi^2  \bigg ] + \xi (\Omega \tau)^2 \bigg \}.
                          \eea
Substituting this expression into Eq. (37) and using Eqs. (1) and (2) we arrive at the result:
% f 59
                         \bea
\frac{\Delta s}{s}  &=& \frac{\alpha^2}{2} \frac{1}{1 + \tilde \sigma^2} \bigg \{1 - \frac{2 \tilde \sigma^2}{1 + \tilde \sigma^2} \bigg [(a^2 - \xi)(\Omega \tau)^2 
   \nn \\ \nn \\ &+ &
\frac{a^2}{2} \bigg (\frac{q}{g} \bigg)^2 
\bigg (\frac{\Delta n}{n}k_F l\bigg)^2
(gR)^2 \Phi^2  \bigg] \bigg \} \, .
                          \eea
 Hence the effect of the field $ \Delta B (r) $ leads to the appearance of an additional term in the expression for the CF conductivity. This term enhances the decrease of the SAW velocity shift upon increasing  the effective magnetic field and thus strengthens the SAW anomaly. The ratio of the magnitudes of the two terms enclosed in square brackets in 
Eq. (59) depends on the magnitude of the density modulation
and on the magnetic field $ B_{eff}. $ When $ B_{eff} $ increases the first term becames larger than the second. Both terms can influence  the CF conductivity and consequently the SAW velocity shift only when the CF-FS is deformed because $ a^2 $ goes to zero for $ \mu = 1 $ (undeformed CF-FS). Hence the deformation of the CF-FS due to the periodic electric field applied to the 2DEG is the most impotant factor determining the 2DEG response to the SAW in the nonlocal regime $ (ql >> 1). $

In metals it is known that the variation in crystalline field by some external effect and/or in electron density, and consequently change of the Fermi energy, can cause a change in the FS topology \cite{26}. Usually these electronic topological transitions represent some changes in the FS connectivity i.e. when a number of voids increases or decreases. Such changes in the FS topology are accompanied by variations in its local
geometry such as forming or losing lines of parabolic points, points of flattening or conic points \cite{21,26}. In the case under consideration the factor which could possibly cause a topological transition in the CF system is increase of the modulating field magnitude. According to the symmetry of the system the transition could occur by means of the formation of two additional small voids arranged along the $ p_y $ direction. At present such a topological change is a conjecture which needs theoretical investigation separately for the case of the composite fermions. 
 
One can expect a change in the CF-FS connectivity to
be accompanied by disappearance of the flattened part on the main void. In this case the anomalous maximum in the SAW velocity shift magnetic  field dependence has to be replaced by a minimum again. Thus assuming  possible relevance of  the electron topological transition   one can explain the disappearance of  the anomalous peak in the SAW velocity shift under increasing the modulation strength. It was mentioned above that this effect was observed in the experiment \cite{12}.

\section{ VI. Summary}

The reference point for this work was the assumption that a periodic electric field applied to the 2DEG in the Quantum Hall Regime at and near one half filling deforms the initially circular CF-FS. We explored possible experimental consequencies of this assumption and found that the unusual peaks arise in the 2DEG response to the SAW propagating in the nonlocal regime over the modulated 2DEG perpendicular to the direction of the modulation. Our results agree with the experiment \cite{12} in the most important points: we explain the anisotropy of the effect, its dependence of the magnitude of the modulating field, its independence of the relation in magnitude between the SAW wave vector $q $ and the modulation wave vector $ g$ and nonlinear (and weak for strong modulation) dependence
of the 2DEG conductivity of $ q.$ All these features originate from the CF-FS deformation (local flattening) due to the modulating field, and exhibit in the region of large $ q $ where small effective parts of the CF-FS strongly contribute to the response of the 2DEG. Our explicit deformed CF-FS model is new, to our knowledge. Again we emphasize that our analysis is carried out in both local and non-local regimes. Since the SAW experiments \cite{6,12} are done in non-local regime it is essential to use the theory for this regime in any interpretation.

 Our work, and that of Refs. \cite{14} and \cite{15} is the assume that the density modulation causes the recently observed magneto-transport anomalies reported in Refs. \cite{12,13}. However our explanation of the SAW anomaly observed in experiments of Ref. \cite{12} substantially differs from that
proposed in these works. The explanation given in Refs. \cite{14} and \cite{15} is based on the extension of the results of their study performed for the longwave limit $ (ql << 1) $ to the region of large $ q $ corresponding to the conditions of the SAW experiments of Ref. \cite{12}. Such an extension
is not justified because it does not account for the important changes in the response of the modulated 2DEG arising in the region of large $q$  due to the CF-FS distortion.

The results obtained show that very slight changes in the shape of the Fermi Surfase for composite fermions -- as well as for the previously studied cases in metal -- can have 
significant effect on magnetotransport in the 2DEG in quantum Hall regime at one half filling. Recent experiments on the SAW propagation above the modulated 2DEG system at
$ \nu = 3/2 $ suggest anomalies similar to that observed at $ \nu = 1/2 $ \cite{27}. We believe that the approach proposed in this work can be applied to analyze the effects due to the CF-FS local geometry inmagnetotransport (especially in the SAW response) to include higher Landau Level partial filling $ (\nu = 3/2, 5/2, $ etc.). To perform this analysis we need to estimate the relative magnitudes of various parameters of the system, first of all the Fermi energy of the CF relevant to the specific half-filled Landau Level (not only the lowest) compared to the energy of the CF interaction with the
modulating field and the CF Fermi wave vector $ q_F $ compared to the wave vector of the periodic modulation $g. $

Additional possibilities for investigations of magnetotransport in modulated 2DEG inFQHE regime at filling factors with even denominators may occur in 2DEG modulated simultaneously in two directions i.e. "cross modulation". Some experiments on the SAW propagation in crossmodulated 2DEG at $ \nu = 1/2 $ were performed recently \cite{27}.

We again remark that our work is based on the charged CF picture for FQHE, for example as derived at $ \nu = 1/2$ in ref.[7,8]. We followed previous work by assuming the CF-FS exists, as supported experimentally and also by a theoretical study \cite{28}. The relevant magnetic symmetry translation
\cite{29}  which would replace Bloch wave vector {\bf k} by a ''good'' quantum index for state labels and transport theory, have not been used to our knowledge. An alternate picture for the FQHE also derived from a Chern-Simons approach \cite{30,31,32}, gives the quasiparticles at $ \nu = 1/2 $ as neutral dipolar objects, with the Hall current being
carried by a set of collective magnetoplasmon oscillators. 
A magnetotransport theory based on this second picture
is proposed recently in Ref. \cite{33}. It is shown there that these two  approaches are equivalent and lead to similar results for observables.

{\it Note added.} In order not to interrupt the continuity in the text, we discuss the solution of Egs. (46) here. The general solution of the system (46) is
   \bea
  \delta v_x & = & C_1 \cos \psi + C_2 \sin \psi
  \nn \\  & - & 
  v_F \frac{\Delta B}{B_{eff}} \int\limits_0^\psi \cos
(gY - gR \cos \psi') d \psi' , 
  \nn \\ \nn \\
 \delta v_y & = & C_1 \sin \psi - C_2 \cos \psi
  \nn \\  & + & 
  v_F \frac{\Delta B}{B_{eff}} \int\limits_0^\psi \cos
(gY - gR \cos \psi') d \psi' , \nn
   \eea 
  where $ C_1 $ and $ C_2 $ arbitrary constants. For the average of $ \delta v_X $ over the unperturbed cyclotron orbit, this gives the results, which differs from our Eq. (47),
  \[
 \big <\delta v_x \big > = v_F \frac{\Delta B}{B_{eff}} \big \{\cos gY J_0 (gR) - \sin g Y J_1 (gR) \big \}.
  \]
   This extra term proportional to $ J_0 (gR) $ arises because the general solution of the system (46) is not periodic in $ \psi. $ The origin of this aperiodicity lies in the system (46), which is valid to first order in the modulation field. When we replaced the coordinate $ "y" $ in the argument of cosine in the expression for $ \Delta n (y) $ by $ Y - R \cos \Omega \tau, $ we arrived at the system of differential equations, which can be reduced to a linear inhomogeneous differential equation with the right side having the same period as the general solution of the corresponding linear nomogeneous differential equation. Such equations describe a resonance that occurs when the frequency of an external perturbation coincides with a natural frequency of free oscillations. Correspondingly, they have to have an aperiodic general solution describing the unbounded increase in the magnitude of oscillations. For our situation it means that strictly we cannot use expansion in powers of a small parameter $ \Delta B /B_{eff} $ to obtain the system (46) because the solution of this system increases in magnitude in time, and we cannot assume it to be small in $ \Delta B $ all the time.
 
The avoid these difficulties, we require the corrections $ \delta V_x $ and $ \delta v_y $ to be periodic in $ \psi , $ which allows us to obtain the expressions (47). Corresponding results of Refs. \cite{17} and {18} were ontained under equivalent assumptions.

\section{acknowledgments}

We thank R.L. Willett and S.H. Simon for discussions and  G.M. Zimbovsky for help with the manuscript. This work was supported in part by a grant from the National Research Council COBASE Program.


\begin{thebibliography}{99}
\bibitem{25} It is shown (see S.H. Simon, Phys.Rev. B {\bf 54}, 13878 (1996)) that the value of $ \sigma_m$ depends on the parameter $ qD$ where $D $ is the distance between the 2DEG and the surface of the sample. The  value used here corresponds to the region of large $ q \ (qD > 1). $
\end{thebibliography}
\end{document}